\documentstyle[multicol,aps,manuscript,psfig]{revtex}
\draft

\begin{document}

\title {An Analytical Study of Coupled Two-State Stochastic
  Resonators} \author{Udo Siewert and Lutz Schimansky--Geier}
\address{Humboldt-Universit\"at zu Berlin, Institut f\"ur Physik, \\ 
  Invalidenstra\ss e 110, D-10115 Berlin, Germany} \date{\today}
\maketitle

\begin{abstract}
  The two-state model of stochastic resonance is extended to a chain
  of coupled two-state elements governed by the dynamics of Glauber's
  stochastic Ising model. Appropriate assumptions on the model
  parameters turn the chain into a prototype system of coupled
  stochastic resonators. In a weak-signal limit analytical expressions
  are derived for the spectral power amplification and the
  signal-to-noise ratio of a two-state element embedded into the
  chain. The effect of the coupling between the elements on both
  quantities is analysed and array-enhanced stochastic resonance is
  established for pure as well as noisy periodic signals.  The
  coupling-induced improvement of the SNR compared to an uncoupled
  element is shown to be limited by a factor four which is only
  reached for vanishing input noise.
\end{abstract}


%\begin{multicols}{2}

\section{Introduction}

The essential point of stochastic resonance \cite{discovery} is the
following: If a stochastic resonator is subjected to an external
influence, commonly referred to as the signal, its response exhibits
signal features, which are most pronounced at a certain level of noise
present in the system. The fascinating aspect of stochastic resonance
is therefore that, counter-intuitively, an increasing noise level does
not steadily deteriorate the transmission of the signal through the
resonator. On the contrary, noise may rather be used to optimise the
transmission process \cite{general}.

The implications of these effects are intensively investigated. In
biological systems, for example, stochastic resonance apparently plays
a role in the neural transmission of information \cite{neural}.  From
a technical point of view one might possibly see the emergence of a
novel type of detector which incorporates an optimal amount of noise
to perform best.  In this regard superconducting quantum interference
devices (SQUIDs) have been studied \cite{Lukens}. They may be made to
detect weak magnetic fields reasonably well without the usual costly
shielding from environmental noise.

In recent years it was shown that the performance of a single
stochastic resonator can be enhanced, if it is embedded into an
ensemble of other stochastic resonators which are properly coupled
\cite{Behn}\cite{neuro}\cite{Morillo}\cite{Lindner}\cite{Marchesoni}:
Compared to being operated isolated the response of the resonator to
the signal increases within the coupled ensemble.  However, if the
coupling becomes too strong this response was found to deteriorate
again. Thus in an ensemble of stochastic resonators the coupling
strength turns out to be a second design parameter: Apart from the
noise level it can be tuned to achieve an optimal performance of the
embedded resonator. This effect was called {\it array-enhanced
  stochastic resonance} \cite{Lindner}.  It will possibly find
technical exploitation and might also be relevant to biological
systems, for example, coupled neurons.

The aim of the present paper is a further investigation of the
phenomena related to array-enhanced stochastic resonance.  To this end
we study a simple prototype system of coupled two-state stochastic
resonators under periodic modulation.  The model we propose allows to
analytically calculate the weak-signal limits of two prominent
stochastic-resonance quantities, spectral power amplification (SPA)
and signal-to-noise ratio (SNR), respectively.  Following the concept
of array-enhanced stochastic resonance both quantities refer to the
response of a resonator as part of the coupled ensemble. The impact of
the coupling on both response measures will be studied.

There is a close link of the present analysis to the work presented in
\cite{Behn} and \cite{Lindner}, which to our knowledge are the most
comprehensive in the field.  However, compared to \cite{Behn}, where
the SPA of a system of globally coupled bistable elements was studied
analytically, too, the present approach offers a more refined
description, since it allows to avoid a mean-field approximation.
Compared to \cite{Lindner}, on the other hand, where the SNR of a
chain of locally coupled elements was investigated in a (rather
extensive) simulation, the SNR can now be obtained analytically. For a
quick review of the results the reader is referred to the 3rd and 4th
passage of the summary section.

From a conceptual point of view the model proposed here can be seen as
an extension of the two-state model developed by McNamara and
Wiesenfeld \cite{MW} to study stochastic resonance in noisy bistable
systems. Instead of considering individual two-state elements as in
\cite{MW} these elements are now arranged in a chain. A simple
next-neighbour interaction is introduced which brings the model close
to the system studied in \cite{Lindner}.  The interaction is chosen in
a way that the resulting evolution of the elements is given by
Glauber's stochastic Ising model \cite{Glauber}.  A detailed
description of this approach is given in the next section which also
provides the necessary background. We note that the connection to the
Glauber model is made for mathematical convenience.  We are not
concerned with the observation of stochastic resonance in Ising
systems which was the central theme in \cite{Neda} and \cite{Brey}.
Our purpose will also lead to the assumption of Arrhenius-type
transition rates of uncoupled unmodulated elements which is unusual in
the context of Ising systems. First results of the present analysis
had been published in \cite{LSG}.

A further interesting approach to arrays of stochastic resonators is c
entered around a response of a more collective nature. Here stochastic
resonance is studied in the summed output of $N$ resonators which need
not necessarily be coupled \cite{Kiss1}\cite{Neiman1}\cite{Collins}.
Recently, it was shown analytically that the SNR of this output
approaches the input SNR, if a sufficiently large number of uncoupled
resonators is used \cite{Neiman2}.  Whether in some way coupling may
still be beneficial for the collective response of the present model
will be the subject of a future study.  Only the summed output of a
very large number of resonators shall be briefly discussed here.  It
will be shown that in this case the SNR always deteriorates under
coupling. This indicates that the coupling-induced improvement of the
performance of stochastic resonators associated with array-enhanced
stochastic resonance is a local rather than a global effect.

\section{From Single to Coupled Two-State Stochastic Resonators}
\label{model}

An early theory of stochastic resonance in noisy bistable systems with
one (generally continuous) variable was worked out by McNamara and
Wiesenfeld \cite{MW}. They studied the effect of a periodic modulation
of these systems in a way which is independent of the precise dynamics
involved. To this end the behaviour of the bistable system was
approximated by a random telegraph process. For simplicity this
process was taken to be symmetric, randomly switching between two
states $\sigma = \pm c$. For convenience we shall consider here $c =
1$.  The probabilities $p(\sigma)$ to find the process in state
$\sigma$ satisfy $p(\sigma) + p(-\sigma) = 1$ and their time-evolution
is governed by
\begin{equation}
\label{MWmaster}
\dot{p}(\sigma) = - \dot{p}(-\sigma) 
                      =  W(-\sigma)p(-\sigma) -  W(\sigma)p(\sigma),
\end{equation}

\noindent
where $W(\sigma)$ denotes the rate of the transition $\sigma
\rightarrow -\sigma$.  These rates must be extracted from the precise
dynamics at hand. To build a theory of stochastic resonance they were
assumed to be approximately given by \cite{MW}

\begin{equation}
 \label{MWrate}
   W(\sigma) = \frac{\alpha}{2}[1-  \sigma \delta \cos( \omega t + \phi)]
\end{equation}

\noindent
with $\delta$ being the small parameter of the theory.  Obviously, at
this level the periodic modulation of the bistable system with
(angular) frequency $\omega$ and phase $\phi$ is taken into account by
the cosine term, only.

This simple model allows to calculate the long-time limit of the power
spectrum of the random telegraph process averaged over a uniformly
distributed phase $\phi$.  The spectrum consists of a continuous part
and a delta function at modulation frequency $\omega$.  The continuous
part of the spectrum is the Lorentzian of the unperturbed process
times a frequency-dependent prefactor. The latter is close to 1 and
governs the modulation-induced transfer of broad-band power to the
delta peak in the spectrum.

From this spectrum McNamara and Wiesenfeld obtained their central
theoretical result with respect to stochastic resonance: An exact
analytical expression for the signal-to-noise ratio (SNR) of the
two-state process. This SNR is defined as the ratio of the weight of
the delta function to the continuous part of the spectrum at
modulation frequency. In view of the present work we shall neglect the
signal-induced suppression of the continuous part of the spectrum and
consider the linear response approximation of the McNamara-Wiesenfeld
result with respect to the small parameter $\delta$.  Doing so one
finds for the SNR

\begin{equation}
  \label{MWsnr}
R_0 =\frac{\pi}{4}  \alpha \delta^2.
\end{equation}

If the approximation (\ref{MWrate}) is performed on a particular noisy
bistable system subject to a periodic modulation, the parameters
$\alpha$ and $\delta$ become functions of noise strength and
modulation amplitude, respectively. The dependence of the SNR on the
noise intensity can then be studied and the occurrence of stochastic
resonance may be established for that particular system.

As an example the overdamped double-well system was given in
\cite{MW}.  The model equation reads

\begin{equation}
\label{doublewell}
\dot{x} = x - x^3 + A \cos( \omega t +\phi ) + \sqrt{2D} \xi(t)
\end{equation}

\noindent
where $\xi(t)$ is Gaussian white noise with $\langle\xi(t)\rangle=0$
and $\langle\xi(t) \xi(t + \tau)\rangle = \delta( \tau) $. Using a
modified Kramers formula for the transition rates valid for
sufficiently low modulation frequencies McNamara and Wiesenfeld found

\begin{equation}
\label{MWparam}
\alpha = \frac{\sqrt{2}}{\pi} \exp \left(-\frac{1}{4D} \right), 
\makebox[0.3cm]{} \delta = \frac{A}{D}.
\end{equation}

The central idea of the present work is to pass from the single
bistable element to a set of coupled bistable elements by extending
the two-state model of McNamara and Wiesenfeld.  To this end we
consider a chain of two-state elements which for convenience is taken
to be of infinite length. If the elements interact the simple
gain-loss balance (\ref{MWmaster}) has to be modified to describe the
evolution of the set of probabilities $p(\bar{\sigma}, t)$ to find the
chain in a particular configuration
$\bar{\sigma}=(.....,\sigma_{k-1},\sigma_k,\sigma_{k+1},.....)$ at
time $t$.  Introducing a formal operator $F_k$ defined for any
function $f(\sigma_k)$ by $F_k f(\sigma_k) = f(- \sigma_k)$ the new
gain-loss balance reads

\begin{equation}
  \label{master}
  \dot{p}(\bar{\sigma}) =  \sum_k(F_k -1)  W_k(\sigma_k)  p(\bar{\sigma}).
\end{equation}

\noindent
With the initial condition $p(\bar{\sigma}, t_0) =
\delta_{\bar{\sigma} \bar{\sigma}_0 }$ the $p(\bar{\sigma}, t)$ are
interpreted as the transition probabilities $p(\bar{\sigma}, t|
\bar{\sigma}_0, t_0)$ of a Markovian process with infinitely many
discrete components. The statistics of the process is fully determined
by (\ref{master}).

To introduce interactions we assume that the transition rates
(\ref{MWrate}) of an element depend on the states of its next
neighbours. A simple choice for this coupling controlled by a
parameter $\gamma$ is

\begin{equation}
  \label{transrate}
   W_i(\sigma_i) =  W(\sigma_i)
\left[ 1 -  \frac{\gamma}{2} (\sigma_{i-1} + \sigma_{i+1})\sigma_i\right],
\end{equation}

\noindent
where $ W(\sigma_i)$ is the McNamara-Wiesenfeld rate (\ref{MWrate}).
With positive $\gamma$ neighbouring elements prefer to be in the same
state, whereas they tend to be in opposite states, if $\gamma$ is
negative.  Both tendencies grow with growing coupling strength
$|\gamma|$. To avoid negative transition rates $|\gamma| \le 1$ has to
be required. We note that the assumed type of coupling is thus able to
model ferromagnetic-type interactions for which a coupling-induced
improvement of the performance of stochastic resonators was found in
\cite{Behn} and \cite{Lindner}. For positive $\gamma$ the present
model is particularly close to the model simulated in \cite{Lindner},
since both are linear arrays of next-neighbour-coupled elements. For
convenience we shall often use the terms ferromagnetic or
antiferromagnetic coupling instead of coupling with positive or
negative $\gamma$.

Of course, other types of transition rates (\ref{transrate}) could be
considered as well.  The advantage of the present choice is that most
of the relevant stochastic properties of the resulting model are
already known. They were studied by Glauber who introduced this model
as a stochastic form of the Ising model \cite{Glauber}. (We note that
in the Glauber model the term $\delta \cos( \omega t + \phi)$ in
(\ref{transrate}) is replaced by a general time-dependent parameter
$\beta$.)  As in the McNamara-Wiesenfeld model it is again possible to
find an analytical expression for the SNR in leading order of the
modulation parameter $\delta$ which is the subject of the next
section. The respective analysis is essentially an exploitation of
Glauber's work.

It may be possible to reduce the dynamics of some coupled systems to
the two-state model given by (\ref{master}) and (\ref{transrate}) and
hence to directly employ the SNR formula derived in the next section.
However, this approach is certainly more involved than the rate
expansion (\ref{MWrate}) needed to make use of the McNamara-Wiesenfeld
SNR.  We shall not pursue this more general aspect of the model in
further detail.  Instead we are interested in devising a simple
prototype system which allows to study the impact of the coupling on
the stochastic-resonance effect.  To this end we are looking for
simple assumptions on the dependence of the model parameter $\alpha$,
$\gamma$, and $\delta$ on some noise intensity and signal amplitude,
respectively. We proceed in two steps:

First, we retain the Glauber's original relation of the present model
to the Ising model given by the Hamiltonian

\begin{equation}
  \label{hamiltonian}
  { \cal{H} } = -J \sum_k \sigma_k \sigma_{k+1}  - \mu H \sum_k \sigma_k.
\end{equation}

\noindent
This relation is defined by an assumption on the probability
$p(\sigma_k)$ to find the $k$-th element in state $\sigma_k$, if all
other elements are fixed: In an adiabatic limit the $p(\sigma_k)$ of
both models are to be identical. This implies that

\begin{equation}
  \label{detailedbal}
\frac{p(\sigma_k)}{p(-\sigma_k)} =
   \frac{W_k(-\sigma_k)}{W_k(\sigma_k)}=
   \exp \left(-\frac{{\cal H}(\sigma_k) - {\cal H}(-\sigma_k)}{kT} \right).
\end{equation}

\noindent
holds which allows to establish a relation between the parameters of
both models. One finds $\gamma = \tanh(2J/kT)$ and $\delta \cos(
\omega t + \phi) = \tanh(\mu H/kT)$, respectively \cite{Glauber}.  The
role of the noise intensity is thus played by the temperature $T$. To
skip unnecessary parameters we set $k=1$ and $\mu=1$.  In the present
paper we assume that $H$, which shall be referred to as the signal, is
given by

\begin{equation}
  \label{forcing}
  H = H_0 \cos (\omega t + \phi).
\end{equation}

\noindent
For weak amplitudes $H_0 \ll T$ this results in $\delta = H_0/T$.  (In
section \ref{srsnr} we shall also consider periodic signals
(\ref{forcing}) with an additional noisy background.)

Second, a plausible assumption has to be made on the temperature
dependence of the time-scale parameter $\alpha $ which is not effected
by requiring (\ref{detailedbal}). It is assumed that it retains the
qualitative dependence (\ref{MWparam}) found by McNamara and
Wiesenfeld for the double-well system, i.e., $\alpha = \alpha_0
\exp(-1/T)$ with properly scaled temperature $T$. This dependence is
typical for many rate processes \cite{H"anggi}. An appropriate
temperature dependence of $\alpha$ is also necessary to turn the
elements into stochastic resonators for vanishing coupling.  For
simplicity $\alpha_0=1$ is considered.

The prototype system we are looking for is thus specified by
(\ref{master}) and (\ref{transrate}) together with

\begin{equation}
\label{params}
  \alpha = \exp \left(-\frac{1}{T} \right), \makebox[0.3cm]{}
  \gamma = \tanh \left(\frac{2J}{T} \right), \makebox[0.3cm]{}
  \delta = \frac{H_0}{T}.
\end{equation}

\noindent
We emphasise that this system is not an approximation of the Ising
system (\ref{hamiltonian}) in the same way as the double-well system
(\ref{doublewell}) is approximated by (\ref{MWmaster}),(\ref{MWrate}),
and (\ref{MWparam}), respectively: The Ising system and the stochastic
Glauber model are based on totally different dynamical concepts.
Nevertheless the devised prototype system appears to capture in a
simple way the essence of the effect of coupling and forcing on
coupled continuous-variable systems like the double well.

To motivate this, let $\triangle U(\sigma)$ denote the height of the
barrier to be surmounted in order to escape from state $\sigma$. In a
very simple approximation the effect of next-neighbour coupling and
forcing on the barrier height could be modelled as

\begin{equation}
  \label{deltaU}
\triangle U(\sigma_k) = 
\triangle U_0 + \triangle U_{add}(\sigma_k),
\end{equation}
\[
\triangle U_{add}(\sigma_k) = J \sigma_k( \sigma_{k+1} + \sigma_{k-1})
+ H_0 \sigma_k\cos(\omega t + \phi).
\]

\noindent
Comparing this to (\ref{hamiltonian}) one finds that within this
simple picture the effect of coupling and forcing on the barrier
heights is the same as their effect on the energy levels of the Ising
model.  (We note that for coupled double-well systems the chosen
approximation would only hold for weak coupling. For strong coupling
the systems may no longer be bistable and the entire concept of a
simple-minded two-state approximation breaks down.)

Within an adiabatic limit and with some noise strength $T$ the
transition rate $W(\sigma_k)$ of the barrier system would roughly read

\begin{equation}
W(\sigma_k) =  \exp \left(- \frac{\triangle U_0}{T} \right) 
\exp \left(- \frac{\triangle U_{add}(\sigma_k)}{T} \right).
\end{equation}

\noindent
From here rates of the form (\ref{transrate}) together with parameters
(\ref{params}) can be obtained in two steps. First, the factor related
to $\triangle U_0$ is kept as the parameter $\alpha$ with a respective
choice of $\triangle U_0$. Second, the factor related to $\triangle
U_{add}$, which describes the effect of forcing and coupling, is
replaced by the respective term of (\ref{transrate}), which is easier
to handle. A connection between both types of rates is again made by
requiring that they lead to identical stationary distributions in the
sense of (\ref{detailedbal}).

A particular close relation between the Glauber-dynamics model and the
barrier system should thus arise within the regime of quasistatic
response. Nevertheless we shall not restrict the following
investigation of the simple prototype system to the quasistatic
regime. Whether or not the suggested model provides insights which are
also relevant to more realistic settings and possibly helpful in the
design of coupled stochastic resonators remains to be seen.

\section{Analytical Solutions}

The aim of this section is to calculate the SNR of the response of a
two-state element embedded into the chain of coupled elements given by
(\ref{master}) and (\ref{transrate}).  This calculation will be done
in terms of the parameters $\alpha$, $\gamma$, and $\delta$ and is
essentially based on Glauber's work. We shall only take account of
terms in leading order of the modulation amplitude $\delta$. The
resulting SNR will then be an extension of the simplified
McNamara-Wiesenfeld result (\ref{MWsnr}).  At the end of the section
the general result will be applied to the prototype system specified
by (\ref{params}).  In addition to the SNR we shall also give an
analytical expression for the so-called spectral power amplification
(SPA).

Following Glauber we start by calculating the averages $\langle
\sigma_k(t) \rangle $.  From (\ref{master}) one derives $ d \langle
\sigma_k \rangle /dt= -2 \langle \sigma_k W_k(\sigma_k) \rangle $. It
results in

\begin{eqnarray}
  \label{mean}
\frac{d \langle \sigma_k \rangle}{d( \alpha t)} &=& - \langle \sigma_k \rangle 
+\frac{ \gamma}{2} \left[ \langle \sigma_{k-1} \rangle + \langle \sigma_{k+1}
 \rangle \right] \nonumber \\
& & +  \left[1 - \frac{\gamma}{2} (r_{k-1,k} + r_{k,k+1})\right]
\delta \cos (\omega t + \phi)
\end{eqnarray}

\noindent
with $r_{i,j}(t) = \langle \sigma_i(t) \sigma_j(t) \rangle$. A closed
set of equations for the $\langle \sigma_k(t) \rangle $ is found by
linearising (\ref{mean}) in $\delta$.  In this case the $r_{i,j}$ can
be taken from the unperturbed model ($\delta = 0$).  After long times
these $r_{i,j}$ read $r_{i,j}= \eta^{|i-j|}$ with $\eta =
\gamma^{-1}(1-\sqrt{1-\gamma^2})$ \cite{Glauber}.

The resulting long-time-limit set of equations can be further
simplified.  Since all elements are forced uniformly, a particular
element cannot be distinguished from any other, once the initial
distribution has been forgotten.  Hence, all elements will have
identical statistics and the indices in (\ref{mean}) can be skipped.
Together with the previous assumption one obtains

\begin{equation}
  \label{simplemean}
\frac{d \langle \sigma \rangle}{ dt}  = -\alpha (1-\gamma) \langle \sigma \rangle  + \alpha  \sqrt{1-\gamma^2} \delta \cos(\omega t + \phi).
\end{equation} 

Obviously, the long-time dynamics of the average $\langle \sigma
\rangle$ is identical to the long-time dynamics of the average of an
uncoupled element with rescaled relaxation rate $\alpha (1-\gamma)$
and rescaled modulation term $\alpha \sqrt{1-\gamma^2} \delta$.  The
resulting long-time limit of the averaged state simply reads
\begin{equation}
\label{meanresult}
\langle \sigma(t) \rangle = q \cos \left( \omega t +\phi +\psi \right), 
\end{equation}
\[
q = q_s \left( 1 + \frac{\omega^2}{\alpha^2(1-\gamma)^2}
\right)^{-\frac{1}{2}}, \makebox[0.3cm]{} \tan \psi = -
\frac{\omega}{\alpha(1-\gamma)},
\]

\noindent
where $q_s$ is the response to static signals $(\omega = 0)$:

\begin{equation}
 \label{meanstatic}
q_s = \delta \sqrt { \frac{ 1+\gamma}{ 1-\gamma} }.
\end{equation}

\noindent
Formally this static response might even exceed $q_s =1$. We conclude
that in this case $\delta$ is not sufficiently small and the
linearisation of (\ref{mean}) is no longer justified. Later on it will
be shown that this does not amount to a considerable restriction.

The power spectrum of an element is determined from the average $
\langle \sigma_k(t) \sigma_k(t+\tau) \rangle$. Like in \cite{MW} its
$t$-dependence is removed by averaging $ \langle .\rangle_{\phi}$ over
an uniformly distributed initial phase $\phi$ of the modulation term
in (\ref{transrate}).  Introducing the correlation function
$c_{kk}(t,\tau) = \langle [\sigma_k(t)- \langle \sigma_k(t) \rangle ]
[\sigma_k(t+\tau)-\langle \sigma_k(t+\tau) \rangle] \rangle $ one
finds

\begin{equation}
 \label{correl1}
\langle \langle \sigma_k(t)  \sigma_k(t+\tau) \rangle \rangle_{\phi} =
 \langle c_{kk}(t,\tau) \rangle_{\phi} +\frac{q^2}{2} \cos(\omega \tau).
\end{equation}

\noindent
The second term on the r.h.s. contributes a delta function with weight
$\pi q^2$ at signal frequency $\omega$ to the (one-sided) power
spectrum as defined below. The first term, on the other hand, forms
the continuous part of the spectrum.

For the purpose of calculating the SNR in leading order of the
modulation amplitude $\delta$ it is sufficient to approximate this
continuous part by the power spectrum of the unperturbed model. This
implies that we neglect any possible signal-induced transformation of
the continuous part of the spectrum as we did to obtain the simplified
McNamara-Wiesenfeld result (\ref{MWsnr}).  For the unperturbed model
the long-time correlation function $c(\tau)$ is not effected by the
average $ \langle .\rangle_{\phi}$. It also no longer depends on the
index of the element. According to Glauber $c(\tau)$ reads

\begin{equation}
  \label{correl2}
c(\tau)  =
 e^{-\alpha |\tau|} \sum_{n=-\infty}^{+\infty} \eta^{|n|}
I_n(\alpha \gamma |\tau|)
\end{equation}

\noindent
with modified Bessel functions $I_n(\alpha \gamma |\tau|)$.  Using the
relation \cite{Gradshteyn}

\begin{equation}
  \label{gradsteyn}
  \int\limits_0^{+\infty} e^{-a x} I_{\nu}(b x) dx = \frac{1}{\sqrt{a^2 - b^2}}
\left( \frac{b}{a+\sqrt{a^2 - b^2}} \right)^{\nu}
\end{equation}

\noindent
which holds for $Re(\nu) > -1$ and $Re(a) >|Re(b)|$ the power spectrum
$ s(\Omega) = 2 \int_0^{+\infty} \cos (\Omega \tau) c(\tau) d \tau $
is calculated.  For the one-sided spectrum defined by $S(\Omega) =
s(\Omega)+s(-\Omega)$ at $\Omega > 0$ we eventually obtain

\begin{equation}
  \label{spectrum}
  S(\Omega) = 4 Re \left(\frac{1+\eta s_2}{s_1(1-\eta s_2)} \right)
\end{equation}

\noindent
with $ s_1 = \sqrt{ (\alpha + i \Omega)^2 - (\alpha \gamma)^2}$ and
$s_2= \alpha \gamma(\alpha +i \Omega + s_1)^{-1}$.

Now the SNR of the response of a two-state element embedded into the
chain can be calculated, which is the central result of this section.
It reads

\begin{equation}
  R^* = \frac{\pi q^2}{S(\omega)}
\end{equation}

\noindent
For vanishing coupling it reduces to the simplified
McNamara-Wiesenfeld result (\ref{MWsnr}).

In addition to the effect of the coupling on the SNR we also wish to
study its impact on the delta peak in the power spectrum. In this
regard the spectral power amplification (SPA) is a convenient measure
(cf. e.g. Jung in \cite{general}).  It is defined as the ratio of
power contained in the signal peaks of output to input spectrum, i.e.,
the ratio of the weights of the respective delta functions.  Because
of its dependence on the input spectrum the SPA depends on the precise
dynamics to be modelled by the two-state chain.

For the prototype system specified by (\ref{params}) the SPA reads
$\rho = (q/H_0)^2$, or explicitly

\begin{equation}
  \label{spa}
\rho = \rho_s
\left( 1 + \frac{\omega^2 \exp\left(\frac{2}{T} \right)}
{\left(1-\tanh\left(\frac{2 J}{T}\right)\right)^2}\right)^{-1},
\end{equation}

\begin{equation}
  \label{spa_static}
\rho_s = \frac{1}{T^{2}} \exp\left(\frac{4 J}{T}\right), 
\end{equation}

\noindent
where $\rho_s$ is the SPA of static signals.  Obviously, the SPA does
no longer depend on the signal amplitude $H_0$.  For convenience we
also remove the $H_0$-dependence of the SNR by considering

\begin{equation}
  \label{snr}
  R = \frac{\pi q^2}{H_0^2 S(\omega)}
     = \frac{\pi \rho}{ S(\omega)}.
\end{equation}

\noindent
For the prototype system the parameter $\eta$ involved in the
unperturbed spectrum (\ref{spectrum}) simplifies to $\eta =
\tanh(J/T)$ \cite{Glauber}.

Both quantities, SPA (\ref{spa}) and rescaled SNR (\ref{snr}), only
depend on three parameters: temperature $T$, coupling strength $J$,
and signal frequency $\omega$. Restrictions arise from $H_0 \ll T$,
which lead to a simple expression for $\delta$, as well as from the
linearisation of (\ref{mean}), which results in an upper bound on
$\delta$ as discussed in connection with (\ref{meanstatic}). Together
one finds

\begin{equation}
 \label{range}
H_0 \ll min(T, T \exp(-2 J / T)). 
\end{equation}

\noindent
It implies that finite $H_0$ place a lower bound on $T$ and an upper
bound on $J$. However, these restrictions are rather weak: $H_0$ can
be made arbitrarily small because its size is immaterial within the
present weak-signal approximation.

We are now in a position to study the impact of the coupling on the
response measures SPA and SNR of a single resonator embedded into the
chain. This will be the subject of the following two sections.

\section{Coupling and  Spectral Power Amplification}
\label{srspa}

The SPA (\ref{spa}) has an unique maximum over temperature $T$ and
coupling parameter $J$ for any time-dependent signal (\ref{forcing}).
The maximum SPA is obtained for a frequency-dependent value $J_{max}$
which is always positive. In other words: A properly tuned
ferromagnetic-type coupling yields the best SPA performance of the
two-state resonator element which is embedded into the chain.

For our simple model this maximum can be studied analytically.  The
partial derivatives of the SPA with respect to $T$ and $J$ are found
to vanish at pairs $(T_{max},J_{max})$ given by

\begin{equation}
\label{simple}
 J_{max} =-(T_{max}/4) ln(2 T_{max} - 1)
\end{equation}

\noindent
and

\begin{equation}
\label{simple2}
\omega^2 = \exp\left( -\frac{2}{T_{max}} \right) \frac{(2 T_{max} - 1)^2 }{T_{max} ( 1 - T_{max} )},
\end{equation}

\noindent
respectively.  For $J_{max}$ to be real and finite, $T_{max} > 1/2$
has to hold.  With this restriction (\ref{simple2}) has a unique
solution $1/2 < T_{max} < 1$ for any given $\omega > 0$. The
corresponding $J_{max}$ is found via (\ref{simple}). Within the given
range of $T_{max}$ one can easily check on (\ref{simple}) that
$J_{max}$ is indeed always positive. An inspection of the second
derivatives of the SPA finally reveals that there is a maximum at
$(T_{max},J_{max})$.  Its peak value is implicitly given by

\begin{equation}
\label{spamax}
\rho_{max} = \frac{1-T_{max}}{T_{max}^2(2 T_{max}-1)}.
\nonumber
\end{equation}
 
From (\ref{simple}), (\ref{simple2}), and (\ref{spamax}) the following
features of the maximum can be derived which are shown in Fig.1:
Tuning the signal frequency $\omega$ from very large to vanishing
small values the temperature $T_{max}$ falls from $1$ to $1/2$. At the
same time, the coupling strength $J_{max}$ as well as the peak height
$\rho_{max}$ increase from vanishing small to very large values.  In
addition, an analysis of the curvatures $\rho_{JJ}$ and $\rho_{TT}$,
both expressed in terms of $T_{max}$, shows that the sharpness of the
maximum grows as its height increases. This indicates that to achieve
optimal performance the system parameters $T$ and $J$ have to be tuned
with increasing accuracy as the signal frequency decreases.

Fig.1 also includes a comparison to the SPA of an uncoupled element.
At given $\omega > 0$ this SPA has a maximum located at a temperature
$T_0$, implicitly given by

\begin{equation}
\label{simple3}
\omega^2 = \exp\left( -\frac{2}{T_0} \right) \frac{T_0}{  1 - T_0 } 
\end{equation}

\noindent
on the interval $ 0<T_0<1$, whereby its peak value is found to be
$\rho_0 = (1-T_0)/T_0^2$.  It can be shown from (\ref{simple2}) and
(\ref{simple3}) that $T_{max}$ always exceeds $T_0$. The fact that
$J_{max}$ was found to be never zero also implies that $\rho_{max}$
always exceeds $\rho_0$.  However, looking at Fig.1 it is obvious that
at frequencies $\omega>1 $ both compared quantities of the coupled
element approach those of the uncoupled one and $J_{max}$ approaches
zero. The coupling-induced increase in the SPA is thus vanishingly
small at sufficiently high frequencies of the signal.

This comparison allows to distinguish between two different types of
SPA behaviour: At low frequencies the SPA is enhanced under
ferromagnetic coupling whereas at high frequencies it is basicly not,
although a tiny increase still occurs. Both situations are illustrated
at selected frequencies in Fig.2 and Fig.3, respectively. The low
frequency SPA exactly reproduces the qualitative effect found
analytically by Jung et al. \cite{Behn} in a system of globally
interacting elements. The high frequency behaviour, on the other hand,
where the SPA does practically not increase under coupling, has to our
knowledge not been reported before.

In both cases Fig.2 and Fig.3 clearly demonstrate that at fixed
coupling parameter $J$ stochastic resonance occurs: The SPA has a
maximum over temperature $T$ which, unfortunately, cannot be
established analytically.  Fig.2(bottom) shows that while
ferromagnetic coupling improves the SPA at any fixed temperature $T$
this improvement is lost, if the coupling becomes too strong. There is
thus an optimal coupling strength $J_{opt}(T)$ for every given
temperature, which can even be exactly calculated at any $\omega > 0$.
One finds

\begin{equation}
  \label{optimalJ}
J_{opt}(T) = {\displaystyle \frac {T}{2}}{\rm 
arctanh} \left( \! 1 + {\displaystyle \frac {1}{2}}
{\displaystyle \frac {{ \omega}^{2}}{{ \alpha}^{2}}} - 
{\displaystyle \frac {1}{2}}\,{\displaystyle \frac {{ \omega}\,
}{{ \alpha}}}
\sqrt {4 + {\displaystyle \frac {{ \omega}^{2}}{{ \alpha}^{2}}}} \!\right).
\end{equation}

\noindent
which is always positive. Since the SPA does not have a further
extremum over $J$, it implies that antiferromagnetic coupling always
decreases the SPA. Moreover, $J_{opt}(T)$ increases with growing
temperature T and decreases as the signal frequency grows.  In Fig.2
the SPA at the optimal coupling strength is included (dash-dotted
curves).  In Fig.3(bottom), where the high-frequency SPA is
illustrated, the coupling-induced improvement of the SPA is hardly
detectable and $J_{opt}(T)$ is almost zero.

Considering the SPA of time-independent signals ($\omega = 0$) given
by (\ref{spa_static}) one finds that it has a maximum over temperature
for antiferromagnetic coupling $J<0$, only, whereby it decreases as
the coupling strength $|J|$ grows.  For $J \ge 0$ the SPA increases
with increasing $J$ as well as with decreasing $T$.  It formally
diverges for $T \to 0$ and $J \to \infty$, respectively. In both cases
the weak-signal limit (\ref{range}) breaks down.

In general, the SPA (\ref{spa}) is given by its static value
(\ref{spa_static}) times a dynamical factor $(1 + d^2)^{-1}$
(cf.(\ref{meanresult})).  Here $d$ is the ratio $d = \omega/(\alpha
(1-\gamma))$ of signal frequency to long-time relaxation rate, which
also governs the phase shift $\psi$ in (\ref{meanresult}).  With
growing $d$ the elements gradually loose their ability to follow the
signal: The SPA weakens and the phase shift grows. This effect occurs,
for example, if the signal frequency $\omega$ increases. Subsequently,
the SPA decreases with growing $\omega$.

The impact of $d$ is also responsible for the occurrence of stochastic
resonance and optimal coupling in the SPA.  Here a decrease of the
long-time relaxation rate $\alpha (1-\gamma)$ plays the crucial role:
The elements dynamics slows down as $J$ increases or as $T$ decreases.
Thus this slow-down occurs whenever the static SPA grows. Hence, the
dynamical factor always counteracts the static SPA as $T$ or $J$ is
changed.  Eventually, the increase of the monotonous static SPA is
outperformed by the decrease of the dynamical factor which results
into a maximum of the SPA over $T$ and $J$, respectively. In other
words, stochastic resonance and optimal coupling occur.

As illustrated in Fig.2 the SPA can be seen as a transition between
two limits, the static SPA given by $\rho_s = (q_s/H_0)^2$ ($d\ll 1$,
dashed curves) and $\rho_s/d^2$ ($d\gg 1$, dotted curves),
respectively. Since both limits intersect at $d=1$, i.e. at $\omega =
\alpha (1-\gamma)$, the plots nicely show the well-known approximate
matching of time scales at the SPA peak.  This matching does not only
occur over $T$ (Fig.2, top), but over $J$ (Fig.2, bottom), too. If the
signal frequency is changed, the peaks of the SPA shift whereby all
curves share the static SPA as a limit. This results into the
qualitative frequency dependence found by McNamara and Wiesenfeld for
the delta function part of the spectrum of the double-well system
(\ref{doublewell}).

Roughly speaking, an improvement of the SPA under ferromagnetic
coupling only occurs at $d<1$, where the impact of the static SPA is
not yet outperformed by the dynamical factor. This explains why there
is no improvement at signal frequencies $\omega \gg 1$: Since under
ferromagnetic coupling $\alpha (1-\gamma) \le 1$ holds, $d<1$ cannot
be fulfilled in this case.  At $\omega = 10$ (Fig.3) the SPA is
already within line width of the limit $\rho_s/d^2 = \alpha^2
(1-\gamma^2) (\omega T)^{-2}$, where the impact of the dynamical
factor prevails. This limit is basicly given by the modulation term in
(\ref{simplemean}).  Its non-monotonous temperature dependence is
known to be no longer associated with a matching of time scales
\cite{MW}.

The static response, which is thus the origin of the coupling-induced
enhancement of the SPA, can be interpreted as follows. Without
coupling and for a given fixed signal the two-state elements prefer to
be in the state of lowest energy (\ref{hamiltonian}) or highest
barrier (\ref{deltaU}), respectively.  Since the signal is
homogeneous, this state is the same for all elements. Without a signal
but under ferromagnetic coupling neighbouring elements tend to be in
the same state, too, although they do not favour a particular state
$+1$ or $-1$.  Taking now signal and coupling together, both effects
add up. The tendency to find the elements in the state favoured by the
signal grows and hence $q_s$ and the static SPA grow compared to the
uncoupled case.

With antiferromagnetic coupling, on the other hand, neighbouring
elements prefer to be in opposite states. This counteracts the effect
of the signal and leads to a decrease in the SPA.

Finally, the decrease of the static SPA with growing temperature
directly follows from (\ref{detailedbal}).  There the imbalance in the
distribution of probability between the two states, which dependent on
the view taken either differ in energy or barrier height, decreases as
the system heats up: The mean response $q_s$ weakens due to increasing
fluctuations.

\section{Coupling and Signal-to-Noise Ratio}
\label{srsnr}

The investigation of the SNR presented here basicly relies on
numerical evaluations of (\ref{snr}). An explicit calculation of the
power spectrum (\ref{spectrum}) already yields a rather complicated
expression which does not lend itself to a detailed analytical study.

At finite signal frequencies $\omega > 0$ the SNR displays the same
qualitative dependence on temperature $T$ and coupling parameter $J$
as shown for the SPA in Fig.2 and Fig.3, respectively.  At low
frequencies the SNR is enhanced under ferromagnetic coupling whereas
at high frequencies it is not.  Due to this close similarity we
omitted the respective plots for the SNR.  We note, however, that this
similarity is by no means a trivial result, since the spectrum
(\ref{spectrum}) is itself a non-monotonous function of $T$ and $J$.

The low-frequency SNR qualitatively reproduces the SNR behaviour found
in a chain of next-neighbour-coupled overdamped double-well systems
simulated in \cite{Lindner}.  It is the first analytical confirmation
of this behaviour.  We note that this correspondence occurs although
the simulation in \cite{Lindner} was performed with strong forcing
while the present model is studied within a weak-signal limit.  As for
the SPA, the high-frequency behaviour of the SNR has to our knowledge
not been reported before.

If the elements are not coupled, their SNR (\ref{MWsnr}) does not
depend on the signal frequency. This was already found in \cite{MW},
if the signal-induced reduction of the continuous part of the spectrum
is neglected, as (\ref{MWsnr}) clearly shows.  Under coupling this
frequency independence of the SNR is lost. For ferromagnetic coupling
numerical evaluations of (\ref{snr}) predict a decrease of the SNR
with growing signal frequency. Only in the limit of low and high
frequencies the SNR approaches constant values.  In both cases the SNR
formula (\ref{snr}) simplifies significantly. One finds

\begin{equation}
   \label{snrlim1}
R_{s} = R_0 (1+\gamma)^2 
       = R_0 \left(\tanh  \left(\frac{2J}{T}\right)+1\right)^2,     
\end{equation}

\begin{equation}
   \label{snrlim2}
R_{HF} = R_0 \sqrt{1-\gamma^2}
       = R_0 \cosh^{-1}\left(\frac{2J}{T}\right),
\end{equation}

\noindent
where $R_0$ is the simplified McNamara-Wiesenfeld SNR (\ref{MWsnr}) of
uncoupled elements.  $R_s$ is the static SNR while $R_{HF}$ represents
the leading order term of its high-frequency expansion.  Fig.4
illustrates the typical qualitative behaviour of the SNR as a function
of the signal frequency. At various frequencies it shows a set of SNR
curves over temperature $T$ and coupling parameter $J$, respectively.
The dashed curves represent $R_s$ and $R_{HF}$, respectively.

The apparent fact that the SNR decreases with growing signal frequency
implies that its static value cannot be exceeded at any other signal
frequency. Then it follows immediately from (\ref{snrlim1}) that the
coupling-induced enhancement of the SNR possesses an upper limit: It
cannot be better than a factor four.  This is a rigorous result for
the general chain model and not limited to the prototype system, where
a special choice of the parameters was made.  Due to the simplicity of
the chain model this result may well reflect a limit of more general
nature for the improvement of a weak-signal SNR. Fig.5 shows the
static SNR over temperature $T$.  It differs qualitatively from the
static SPA which according to (\ref{spa_static}) has a monotonous
behaviour.

Turning briefly to antiferromagnetic coupling one finds that the
continuous part of the spectrum is insensitive to the sign of the
coupling parameter $J$. Now even the static SNR decreases under
coupling. As shown in Fig.4(bottom) it is found below the
high-frequency SNR curve, i.e., here the SNR increases with growing
signal frequency.  Since at medium frequencies again a transition
occurs between both curves similar to the transition in Fig.4(top), a
second maximum in the SNR over $T$ may emerge in this situation.

The investigation of the impact of the coupling on the SNR can also be
extended to the more natural situation, where the signal itself is
embedded into noise.  The question to address is, whether the
improvement of the SNR, which was established for independent internal
noise sources, can still be found with additional coherent external
noise. (This problem does not occur for the SPA, where only the height
of the signal peak is of interest.)

To study this case it is assumed that the input spectrum consists of
the previous signal peak described by $\pi H_0^2 \delta(\Omega -
\omega)$ and a noise part $N(\Omega)$.  Within the weak-signal limit
it was shown that the signal peak of the input spectrum leads to a
respective peak in the output spectrum at signal frequency, only: No
additional peaks at multiples of that frequency occur. Therefore, any
additional contribution to the continuous part of the output spectrum
at signal frequency can only arise from $N(\Omega = \omega)$.  In
analogy to (\ref{meanresult}) one finds that this additional
contribution is given by $(q/H_0)^2N(\omega)$.

Together with (\ref{snr}) the resulting SNR $R_{noisy}$ reads, again
in units of $H_0^2$,
\begin{equation}
  \label{noisysnr1}
  R_{noisy}   = \frac{\pi q^2}{H_0^2 S(\omega)+ q^2 N(\omega)}.
\end{equation}
It can be expressed in terms of the SNR $R$ (cf.(\ref{snr})) and the
input SNR $R_{input}=\pi /N(\omega)$ written in units of $H_0^2$,
respectively.  One finds
\begin{equation}
  \label{noisysnr2}
\frac{1}{R_{noisy}}   = \frac{1}{R} + \frac{1}{R_{input}}.
\end{equation}

\noindent
$R_{noisy}$ is thus a steadily growing function of the previously
studied SNR $R$, whereby it cannot exceed the input SNR $R_{input}$.
(The latter is a meanwhile well-known result of linear response theory
\cite{Neiman2} which is in fact the limit we are taking here.)  Since
$R_{input}$ is constant at fixed signal frequency $\omega$, a
coupling-induced increase of $R$ will lead to an increase in
$R_{noisy}$, too.  The improvement of the SNR under coupling is thus
preserved with external coherent noise. From (\ref{noisysnr2}) one can
easily show that the maximum enhancement of $R_{noisy}$ compared to
the uncoupled chain reaches the previously found factor four for
vanishing input noise, only.  In general, this factor is smaller
approaching 1, if the input noise is so strong that the input SNR and
hence $R_{noisy}$ go to zero.

At the end we shall make a brief excursion to the collective response
of the chain, which was mentioned in the introduction. We consider the
following sum which involves the states of $N$ elements:

\begin{equation}
\label{global0}
M(t)= \sum_i^N \sigma_i(t).
\end{equation}

\noindent
It is not difficult to show that the SPA of this new quantity is
simply $N^2$ times the previously studied SPA (\ref{spa}). For the
SNR, however, this new situation is completely different. Previously
the continuous part of the spectrum was determined from the
auto-correlation function $c_{kk}(t,\tau)$ of an element alone (cf.
(\ref{correl1})). Now this spectrum will involve cross-correlation
contributions $c_{jk}(t,\tau)$ of different elements $j \ne k$, too.

The new contributions to the continuous part of the spectrum change
the qualitative behaviour of the SNR under coupling. We shall
demonstrate this in the limit $N \to \infty$. For this case Glauber
calculated the unperturbed spectrum \cite{Glauber} which reads in its
one-sided version

\begin{equation}
\label{global3}
S_{M}( \Omega ) = 
N \frac{ 4\alpha \sqrt{1-\gamma^2}}{\alpha^2 ( 1 -
  \gamma)^2 + \Omega^2}.
\end{equation}

\noindent
Inserting this spectrum into (\ref{snr}), multiplied by $N^2$ due to
the mentioned increase of the SPA, one finds for the new SNR per
element
\begin{equation}
\label{global5}
R_M = \frac{\pi}{4} \alpha  \sqrt{1 - \gamma^2} \delta^2,
\end{equation}

\noindent
which, of course, again reduces to the simplified McNamara-Wiesenfeld
SNR (\ref{MWsnr}) for vanishing coupling.

Clearly, the SNR is now a decreasing function of the coupling
parameter and the sign of the latter is no longer important. Hence,
there is a drastic difference in the SNR of single and collective
response with respect to coupling. Of course, we have so far merely
investigated the limits of (\ref{global0}), $N=1$ and $N \to \infty$,
respectively. We expect that the SNR of the collective response of
only a few elements still increases under coupling \cite{Neiman1}. A
detailed analysis of the present model with respect to these
collective effects will be the subject of a further investigation.

\section{Summary}
\label{sum}

In this paper we extended the two-state model of stochastic resonance
introduced by McNamara and Wiesenfeld to a chain of infinitely many
coupled two-state elements which are periodically modulated. The
interaction of the elements was chosen in a way that the chain evolves
according to Glauber's stochastic Ising model.  In analogy to the work
of McNamara and Wiesenfeld and based Glauber's results analytical
expressions for the signal-to-noise ratio (SNR) and the spectral power
amplification (SPA) have been obtained in the limit of weak
modulations.  Here both quantities refer to the response of a single
element as part of the chain.

Instead of approximating the dynamics of a particular coupled system
by the chain model, we used the latter to build a prototype system
which hopefully captures the essential features of an entire class of
coupled stochastic resonators.  To this end additional assumptions
were made on the dependence of the model parameters on some noise
intensity and signal amplitude, respectively. The prototype system was
used to study the effect of the coupling on the response of a single
resonator embedded into the chain.

The results show that array-enhanced stochastic resonance occurs for
ferromagnetic-type coupling in SPA and SNR. The qualitative features
of the effects reproduce those previously found in coupled stochastic
resonators. The simplicity of the chain model allowed for a detailed
analytical investigation of the SPA.  For the SNR the observed effects
have been confirmed analytically for the first time. In addition, it
was found that the improvement of the single-resonator response
compared to the response of the uncoupled resonator still occurs, if
the signal is embedded into noise. For the SNR this improvement was
shown to be limited by a factor four which is reached for vanishing
input noise, only.  A brief excursion into the collective response of
$N$ resonators, on the other hand, disclosed that coupling cannot
improve the SNR, if $N$ is very large.

A closer look at the mechanisms behind the effects revealed that in
the present model an improvement of the stochastic resonators under
coupling is associated with the regime of quasistatic response.  Since
the model studied here possesses an upper bound to the time scale of
its dynamics, the desired improvement is essentially restricted to
sufficiently slow signals.  For the SPA this improvement is based on a
stronger tendency of the two-state elements to align in parallel, if
signal and ferromagnetic coupling act together, compared to this
tendency caused by the signal alone. The reason why the improvement of
the SPA is lost, if the coupling is too strong, was found to be the
slow-down of the system dynamics under coupling: It simply prevents
the resonators from responding quasistatically.

We would like to thank an anonymous referee whose valuable
recommendations helped to shape the final version of this paper.

%\end{multicols}

\begin{figure}[htbp]
\centerline{\psfig{file=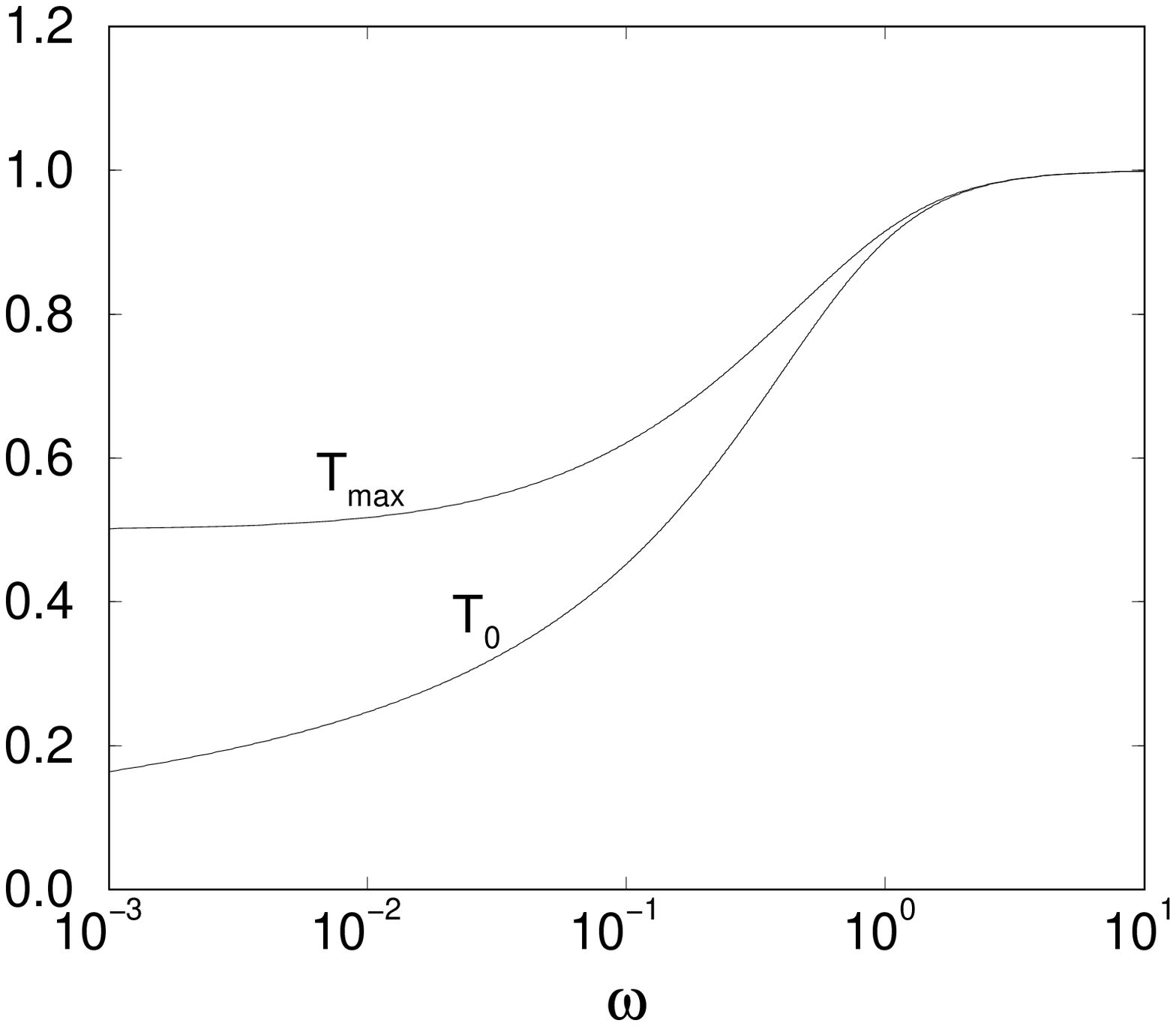,height=8cm}}
\centerline{\psfig{file=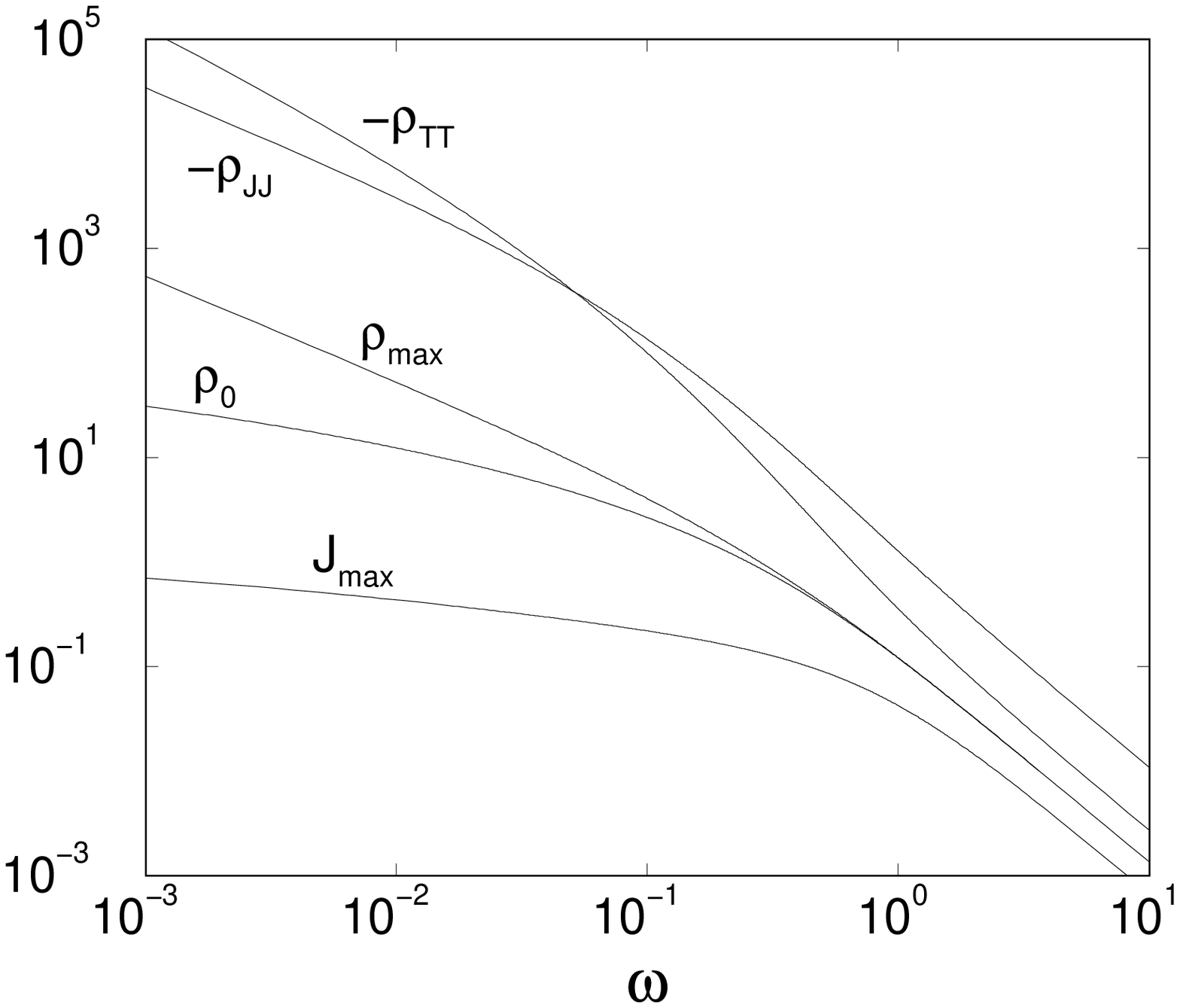,height=8cm}}
\vspace*{1cm}
  \caption{Features of the unique maximum of the SPA $\rho$ over
temperature $T$ and coupling strength $J$ shown in dependence on the
signal frequency $\omega $. Plotted are the position ($T_{max}$,
$J_{max}$) of the maximum, its height $\rho_{max}$, and the sharpness
of the maximum expressed by the curvatures $\rho_{TT}$ and
$\rho_{JJ}$, respectively. In the uncoupled model ($J=0$) the SPA has
a maximum at $T_0$ with a height $\rho_0$.}
  \label{fig:1}
\end{figure}

\begin{figure}[htbp]
\centerline{\psfig{file=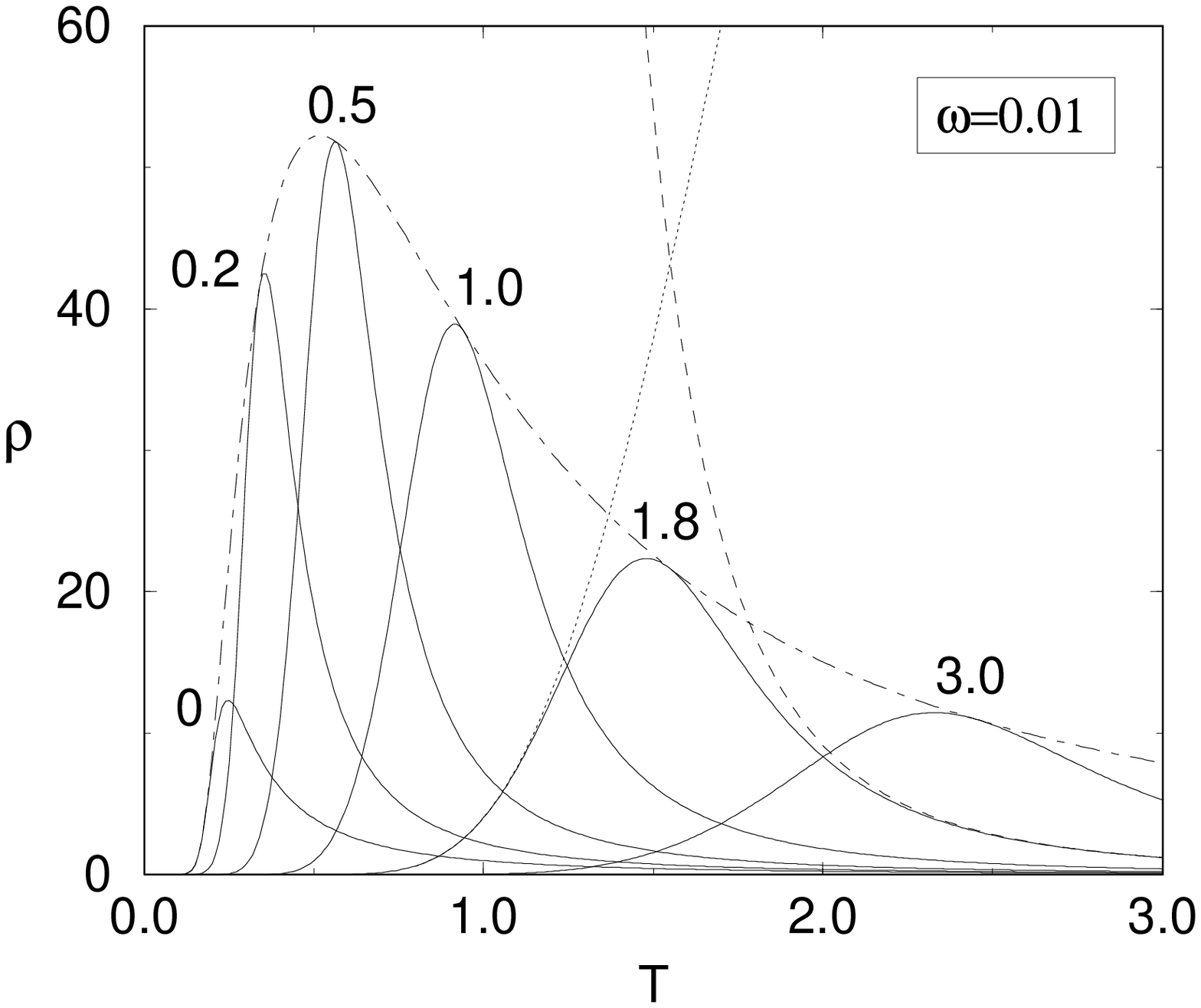,height=8cm}}
\centerline{\psfig{file=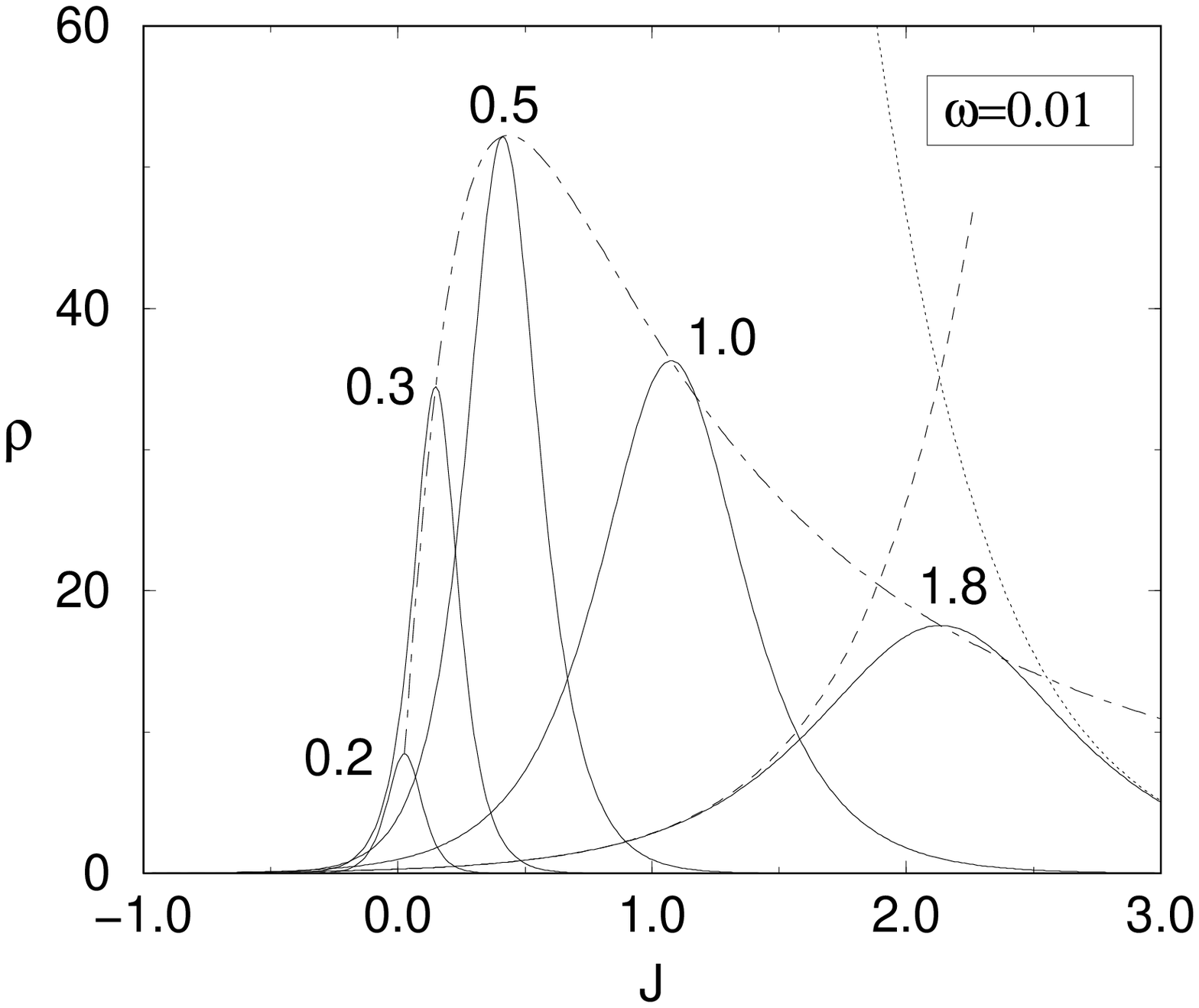,height=8cm}}
\vspace*{1cm}

  \caption{Typical qualitative behaviour of the SPA $\rho$ at low signal
frequencies for various values of coupling strength $J$ (top) and
temperature $T$ (bottom).  Dashed and dotted curves correspond to SPA
limits discussed towards the end of this section. The dash-dotted
curves represent the SPA at optimal coupling.}
  \label{fig:2}
\end{figure}

\begin{figure}[htbp]
\centerline{\psfig{file=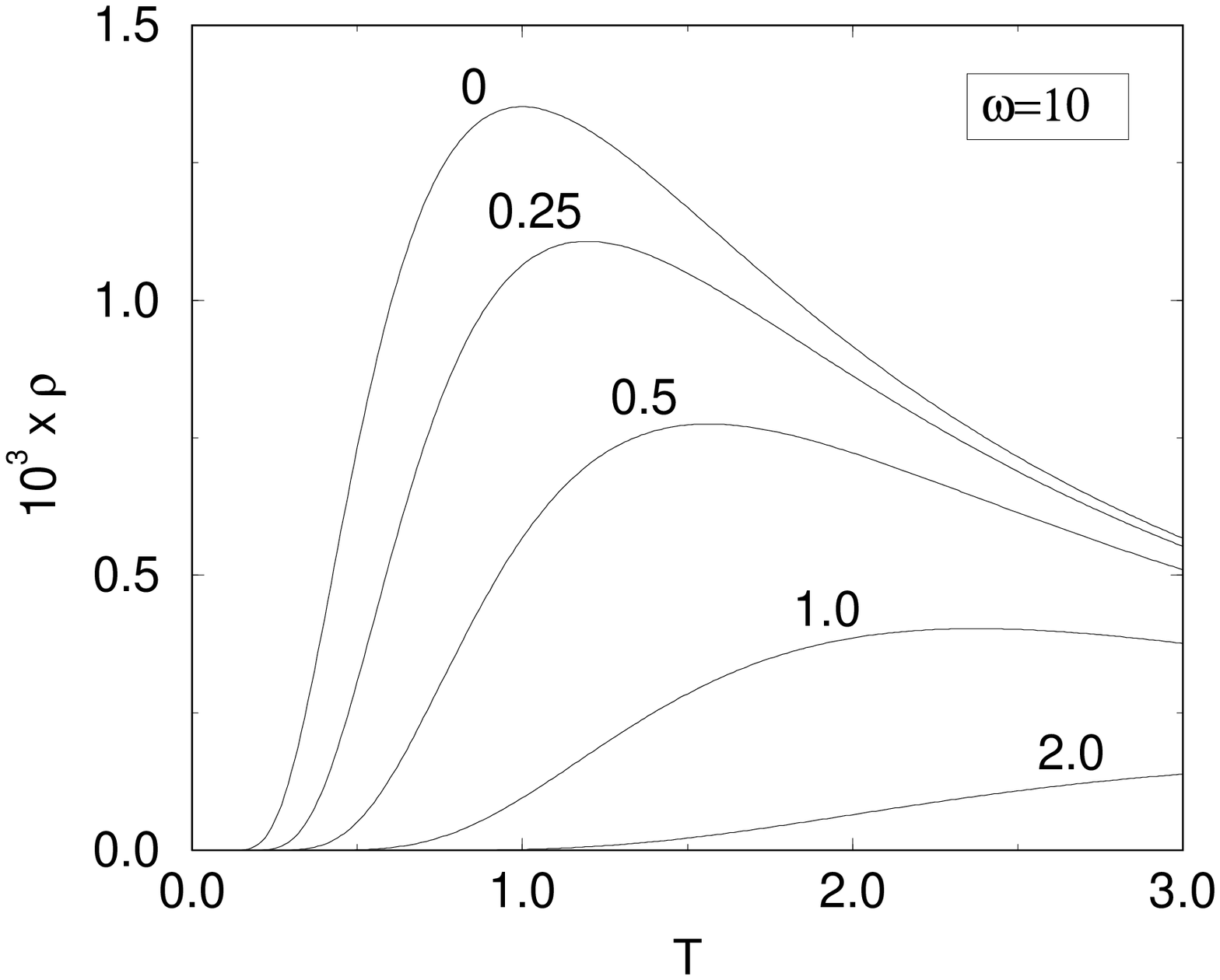,height=8cm}}
\centerline{\psfig{file=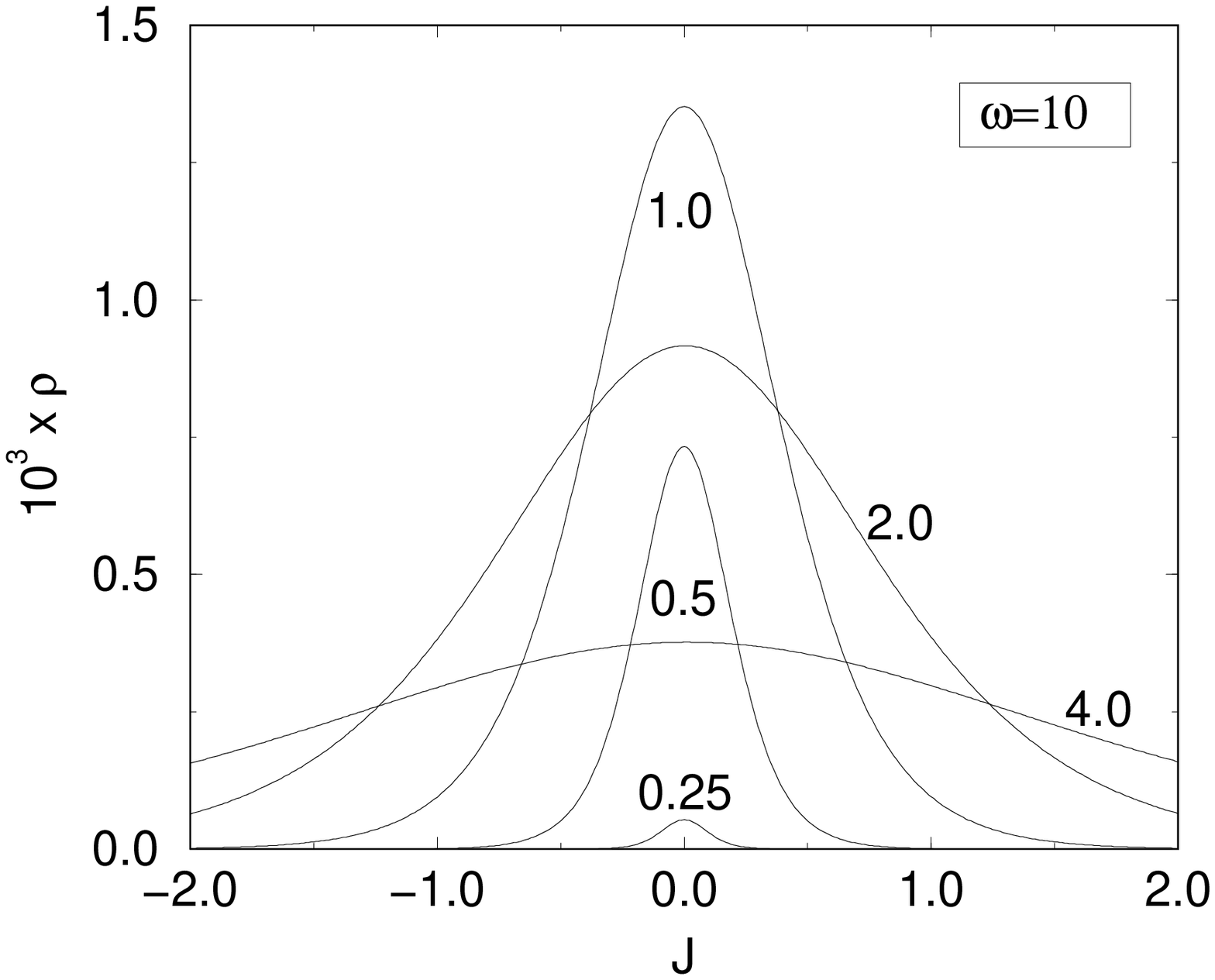,height=8cm}}
\vspace*{1cm}

  \caption{Typical qualitative behaviour of the SPA $\rho$ at high signal
    frequencies for various values of coupling strength $J$ (top) and
temperature $T$ (bottom).}
  \label{fig:3}
\end{figure}

\begin{figure}[htbp]
\centerline{\psfig{file=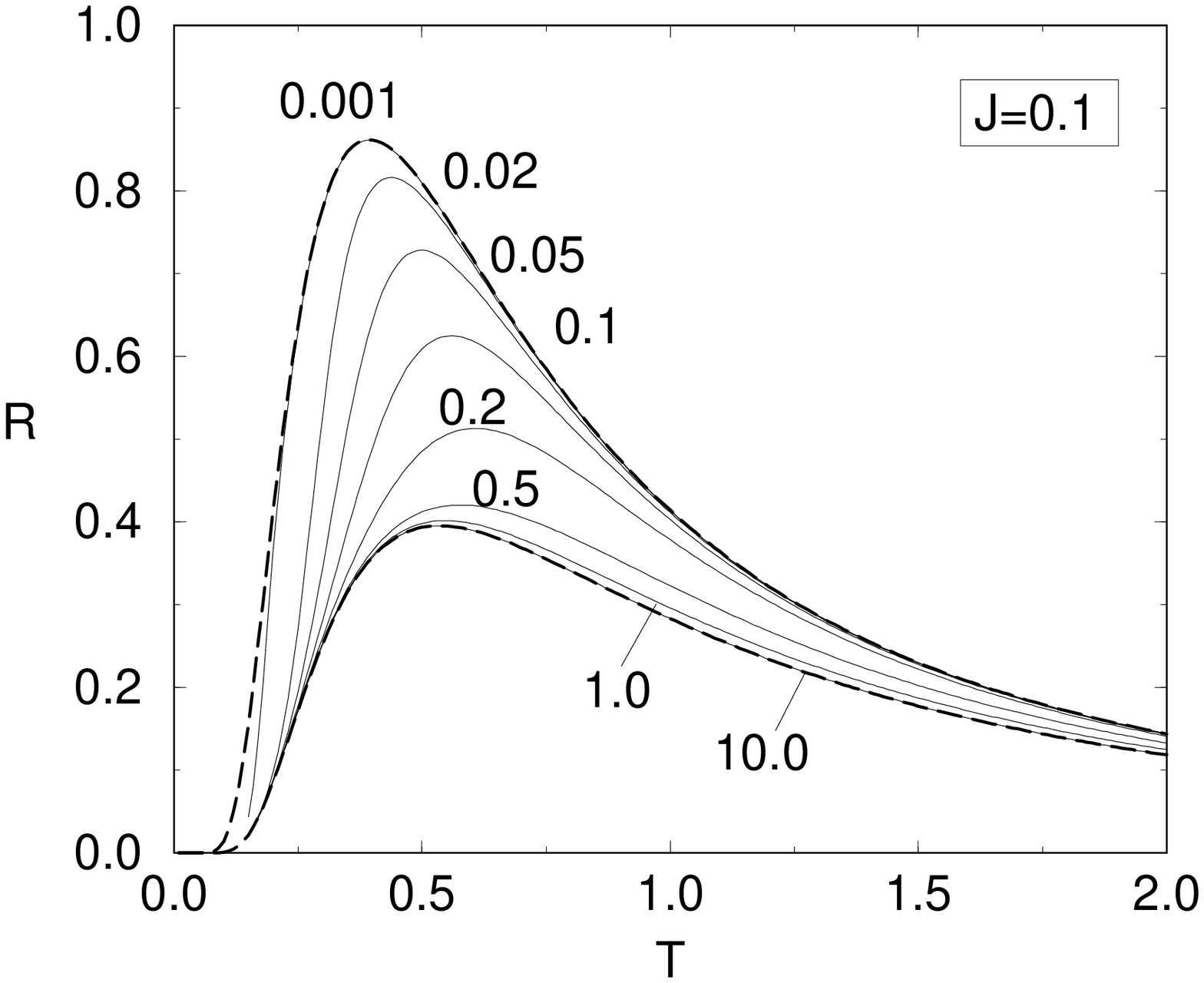,height=8cm}}
\centerline{\psfig{file=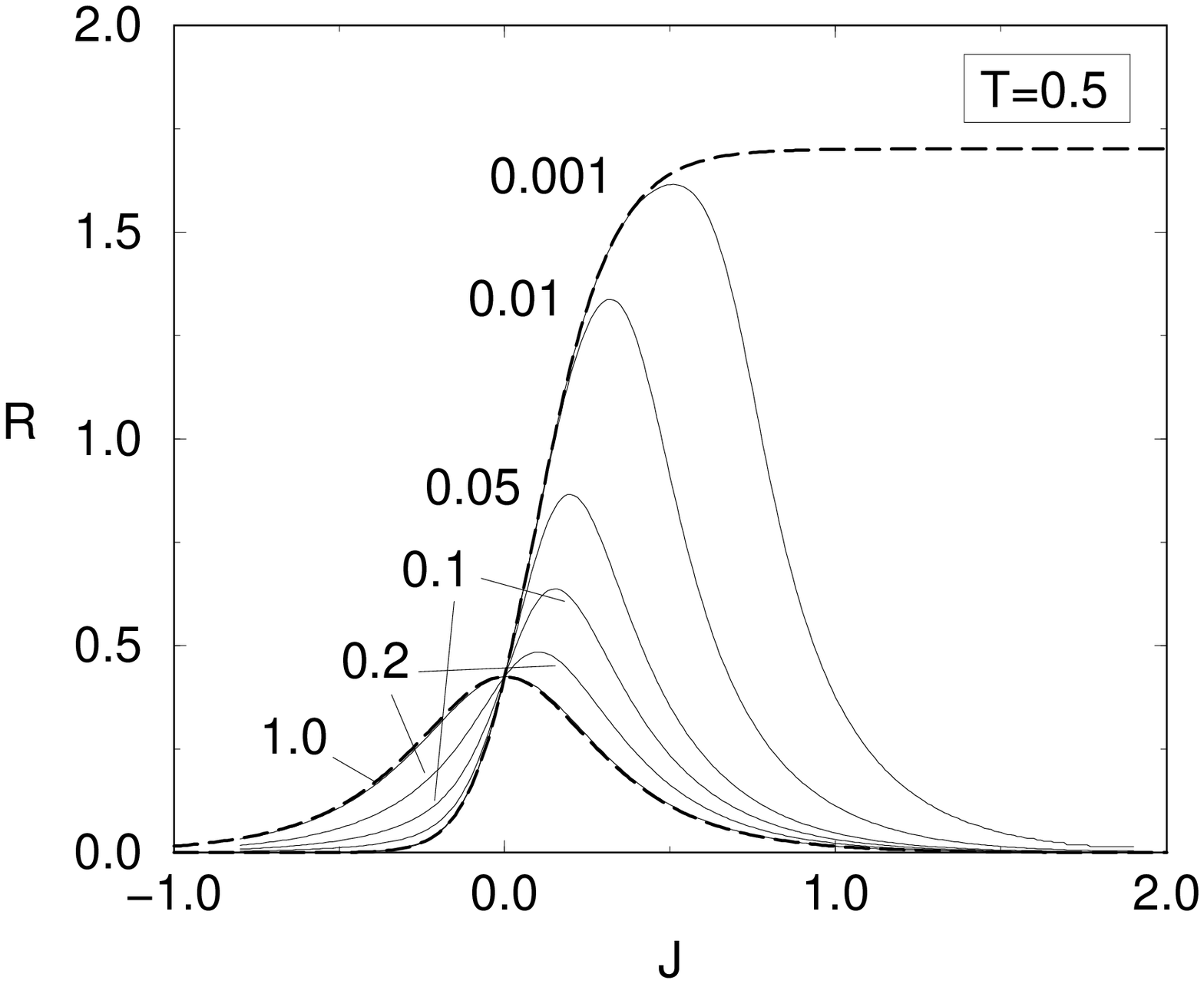,height=8cm}}
\vspace*{1cm}
  \caption{The frequency-dependence of the SNR $R$. The upper dashed
    curves represent the static SNR. The lower dashed curves show the
high-frequency expansion of the SNR.}
\end{figure}

\begin{figure}[htbp]
\centerline{\psfig{file=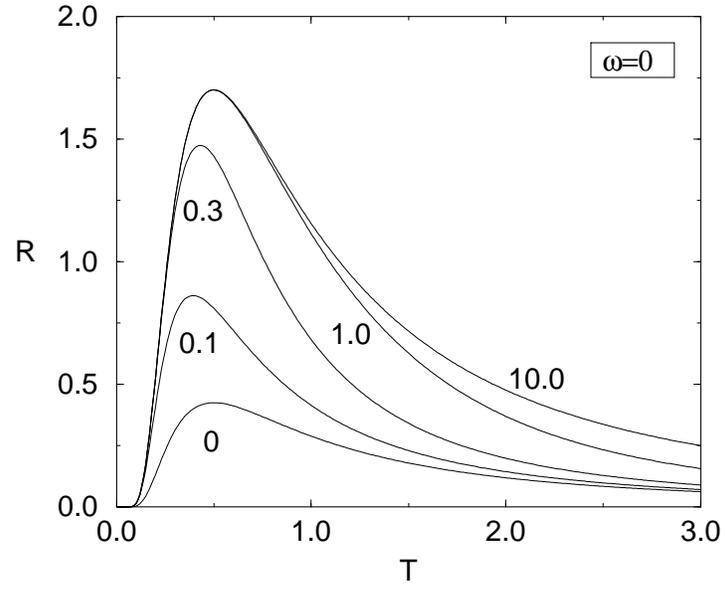,height=8cm}}
\vspace*{1cm}
  \caption{Static SNR $R$ for various values of the coupling parameter
    $J$. The SNR at $J=10$ is within line width of the amplification limit
$4 R_0$.}
  \label{fig:5}
\end{figure}

\end{document}